\documentclass[conference]{IEEEtran}
\IEEEoverridecommandlockouts

\renewcommand{\tablename}{Table}
\renewcommand{\thetable}{\arabic{table}}
\usepackage{lipsum}
\usepackage{array}
\usepackage{caption}
\usepackage{cite}
\usepackage{amsmath,amssymb,amsfonts}
\usepackage{algorithmic}
\usepackage{multirow}
\usepackage{algorithm}

\usepackage{float} 
\usepackage{graphicx}
\usepackage{textcomp}
\usepackage{xcolor}
\usepackage{booktabs} 
\def\BibTeX{{\rm B\kern-.05em{\sc i\kern-.025em b}\kern-.08em
    T\kern-.1667em\lower.7ex\hbox{E}\kern-.125emX}}
\begin{document}

\title{Quantum-Inspired Portfolio Optimization In The QUBO Framework}


\makeatother

\author{
    Ying-Chang Lu\textsuperscript{1},  Chao-Ming Fu\textsuperscript{1*}, Lien-Po Yu\textsuperscript{2}, Yen-Jui Chang\textsuperscript{3,4},  Ching-Ray Chang\textsuperscript{1,4}\\
    \textsuperscript{1}Department of Physics, National Taiwan University, Taipei, Taiwan \\
    \textsuperscript{2}Institute for Information Industry, Taipei, Taiwan \\
    \textsuperscript{3}Master Program in Intelligent Computing and Big Data, Chung Yuan Christian University, Taoyuan City, Taiwan \\
    \textsuperscript{4}Quantum Information Center, Chung Yuan Christian University, Taoyuan City, Taiwan\\
    *Corresponding Author: chaomingfu@phys.ntu.edu.tw
}

\maketitle


\begin{abstract}

A quantum-inspired optimization approach is proposed to study the portfolio optimization aimed at selecting an optimal mix of assets based on the risk-return trade-off to achieve the desired goal in investment. By integrating conventional approaches with quantum-inspired methods for penalty coefficient estimation, this approach enables faster and accurate solutions to portfolio optimization which is validated through experiments using a real-world dataset of quarterly financial data spanning over ten-year period. In addition, the proposed preprocessing method of two-stage search  further enhances the effectiveness of our approach, showing the ability to improve computational efficiency while maintaining solution accuracy through appropriate setting of parameters. This research contributes to the growing body of literature on quantum-inspired techniques in finance, demonstrating its potential as a useful tool for asset allocation and portfolio management.

\end{abstract}


\section{Introduction} 

In the realm of financial investment, portfolio optimization constitutes a pivotal pursuit that seeks to minimize risk for a given level of return or maximize return for a given level of risk by diversifying investment across different asset classes. Pioneered by Harry Markowitz in 1952\cite{markowits1952portfolio}, the Modern Portfolio Theory (MPT) has adapted to accommodate the intricacies of a varied investment environment. The concept of diversification as the core of MPT aids in constructing portfolios that effectively balance risk and return. This balance is often depicted by an efficient frontier—a graphical representation of optimal portfolios which offer the maximum potential expected return for a given level of risk.

Nonetheless, the practical implementation of portfolio optimization encounters numerous challenges, mainly due to the inherent complexity of financial markets and the limitations of classical computing particularly in solving certain type of large and complex problems such as those involving large portfolios with complex real world constraints.

In the field of portfolio optimization, several studies have utilized classical optimization solvers (CPLEX, Gurobi, etc.) to address diverse constraints and explore a range of strategies to optimize performance \cite{jin2016constrained, anagnostopoulos2011mean}; others have applied machine learning techniques to develop effective models to address the challenges of inherently complex real world applications like handling high-dimensional financial data \cite{9350582, chopra2021application, cohen2022algorithmic}. On the other side, in order to tackle the ever-increasing large and complex optimization problems with excessively large search space that are intractable for classical computers, the more advanced and emerging unconventional computing such as quantum computing has extensively researched and developed in recent years \cite{palmer2021quantum, slate2021quantum}. Inspired by quantum computing, the quantum-inspired classical computing methods, particularly in the framework of Quadratic Unconstrained Binary Optimization (QUBO) \cite{10.3389/fphy.2014.00005}, has emerged as a powerful tool for solving computationally hard combinatorial optimization problems that are ubiquitous in many important applications like large-scale portfolio management \cite{grant2021benchmarking, lucas2014ising, glover2018tutorial}.

Besides, a growing class of hardware-accelerated quantum-inspired optimization solvers, such as digital annealer\cite{aramon2019physics} — a specialized application-specific CMOS hardware engineered for solving fully connected QUBO problems, can further advance the exploration ability of the quantum-inspired approach toward solving computationally hard combinatorial optimization problems \cite{rahman2019ising, cohen2020ising, kalehbasti2021ising}.

In the context of optimization in the QUBO framework, the parameter tuning primarily refers to the process of selecting the optimal values of parameters in a QUBO problem formulation, which is challenging but essential to affect the performance of optimization solver being used. There have been a variety of methods being developed to estimate the optimal penalty coefficients of a QUBO problem that is very often converted from an original constrained optimization problem using penalty methods, for instances, exact and sequential methods, or specifically hybridizing exact and sequential methods aiming at a general, automatic way to find valid penalty coefficients \cite{garcia2022exact}; lower bound estimation for the penalty coefficient for equality-constrained minimization problem \cite{VERMA2022100594}. In this paper, we propose a method based on Monte Carlo simulations to efficiently estimate the penalty coefficient in the QUBO formulation. 

Preprocessing techniques are often used for hard optimization problems to reduce the time to find good solutions, and there have been a variety of efficient preprocessing methods being developed for QUBO problem reduction, for instances, slack variable reduction techniques to convert inequality constraints into equality\cite{verma2021variable, VERMA2022100594}; linear constraints of lower and upper bounds in slack variable to cut down the numbers of formulation\cite{sakuler2023real}; a new workflow to solve portfolio optimization problems with more diversified portfolios on annealing platforms \cite{app122312288}. To further address the challenges in efficiently finding high-quality solutions in complex scenarios, a two-stage search algorithm is proposed in that an initial broad (rough) search is first employed to quickly identify feasible solution which is then used by a refined search in the second stage to improve the solution accuracy.

The rest of the paper is organized as follows. Section II introduces the proposed methods. Section III presents the experimental results on a real-world dataset. Section IV discusses the effects of the proposed methods on the performance of portfolio optimization. Section V concludes the paper with future directions.

\section{Method}

\subsection{QUBO Formulation for the Markowitz Portfolio Optimization Problem}

In this study of Markowitz portfolio optimization, an investor aims to distribute investment among \( n \) assets to maximize their portfolio returns while managing the portfolio risk.

Let \( R_p \) denote the expected return from the entire portfolio which can be defined as:

\begin{equation}
\label{expected_return}
R_p = \sum_{i=1}^n x_i r_i
\end{equation}

where \( r_i \) is the expected return from asset $i$ and \( x_i \in R\) are weight of asset \( i \) in the portfolio, respectively, and \( n \) is the total number of assets. The risk, measured as portfolio variance \( \sigma_p^2 \), is given by:

\begin{equation}
\label{risk}
\sigma_p^2 = \sum_{i=1}^n \sum_{j=1}^n \text{Cov}(i, j) \cdot x_i x_j 
\end{equation}

where \( \text{Cov}(i, j) \) is the covariance between assets \( i \) and \( j \). The goal is to find the optimal set of weights \( \{x_i\} \), with requirement of \(\displaystyle\sum_i^n x_i = 1\), that either minimize risk for a given return or maximize return for a given risk by creating an efficient frontier of optimal portfolios.

The Markowitz portfolio optimization problem to study in this work is defined as follows:
    \begin{equation}
    \label{our_method}
        \begin{aligned}
   & \text{minimize} && \sum_{i=1}^n \sum_{j=1}^n \text{Cov}(i,j) \cdot x_i x_j\\
   & \text{subject to} && R_p \geq R
   \end{aligned}
    \end{equation}
where \(R\) is the lower bound on the expected return.

As constrained problems are relatively more complex than unconstrained problems, in order to solve the above constrained problem using the unconstrained optimization methods, it can be converted into an unconstrained problem by introducing the commonly used quadratic penalty function (the penalty terms are the square of the constraint violations), which can be mathematically represented as follows:

\begin{equation}
\label{eq:qubo_MPT_normal}
\begin{aligned}
\mathbf{H} & = \theta \cdot \left(\sum_{i=1}^n \sum_{j=1}^n \text{Cov}(i,j) \cdot x_i x_j\right) \\ & - M \cdot \left(\sum_{i=1}^n x_i r_i - R \right) ^2
\end{aligned}
\end{equation}

where \( \theta \) is a scaling parameter for the original objective term (risk), and $M$ is a penalty parameter for the
constraint term (return) that penalizes constraint violation relative to the objective function.

In the context of quantum-inspired optimization in the QUBO framework, the optimization problem is mapped into a QUBO model as follows:

\begin{equation}
    \label{eq:qubo}
    \mathbf{H} = \sum_{i = 1}^n{a_ix_i} +  \sum_{i=1}^n \sum_{j=1}^n {b_{i,j}x_ix_j} 
\end{equation}

where $\mathbf{H}$ denotes the objective function, $x_i \in \{0, 1\}$ is the i-th binary decision variable, $a_i$'s are  the coefficients of the linear terms, and $b_{i,j}$'s are the coefficients of the quadratic terms.

Moreover, many of the constrained optimization problems, as are often found in real-world cases, can be effectively re-formulated as a QUBO model by introducing penalty function into the objective function as follows:

\begin{equation}
    \label{eq:qubo_constraint}
    \mathbf{H} = \sum_{i=1}^n{a_ix_i} + \sum_{i=1}^n \sum_{j=1}^n{b_{i,j}x_ix_j}  + M \cdot g(x)
\end{equation}

where \(g(x)\) is the exterior penalty function, \(M\) is the penalty coefficient, and the index \(i\) refers to each decision variable in the system.

In order to solve the problem with proportional investments in the QUBO framework, a binarization technique is introduced in the QUBO model via binary expansion to accommodate proportional investments, which as a result leading to the following QUBO formulation\cite{sakuler2023real}:

\begin{equation}
\label{binary_sub}
\begin{aligned}
\mathbf{H} = & \theta\cdot\sum_{i=1}^n \sum_{j=1}^n \text{Cov}(i,j) \left(\sum_{k=1}^K p_k 2^{k-1} x_{i,k}\right) \left(\sum_{k=1}^K p_k 2^{k-1} x_{j,k}\right) \\ & - M \cdot \left[\left(\sum_{i=1}^n  \left(\sum_{k=1}^K p_k 2^{k-1} x_{i,k}\right) \cdot r_i\right)-R\right]^2 
\end{aligned}
\end{equation}

where $x_{i,k}$ denotes the $k$-th binary decision variable associated with asset $i$, representing a fraction of the proportional investment. The term $p_K = 1/2^K$ represents the granularity level of the binary encoding using $K$ bits, facilitating a granular allocation of investment proportions.

\subsection{Selection of Penalty Coefficients in QUBO Formulation}

\subsubsection{Penalty Coefficient Estimation Using Monte Carlo Simulations}

Parameter tuning plays a significant role in optimizing model performance such as the speed and quality of the solution \cite{garcia2022exact, VERMA2022100594, verma2021variable}. Finding optimal values for the parameters of \(\theta\), \(K\), and \(M\) in Equation~\ref{binary_sub} is essential for achieving an optimal balance between the expected return and risk in portfolio optimization. By leveraging insight from~\cite{VERMA2022100594} the technique of calculating the estimate of the lower bound of penalty coefficient $M$ and employing Monte Carlo simulation technique, an estimation method is proposed with its procedures outlined as follows and summarized in Algorithm~\ref{al:monte_carlo_method}:

\begin{itemize}
    \item Employ the Monte Carlo method to generate a set of feasible solutions, say \(x_{\text{set}}\).
    \item Find the feasible solution \(X\) in \(x_{\text{set}}\), say \(x_{\text{from}}\), that minimize \(A x - b\), where \(A\) is an \(n \times n\) matrix and \(b\) represents a vector.
    \item From the remaining feasible solutions in \(x_{\text{set}}\), find the subset, say \(x_{\text{to}_\text{set}}\), that matches the following equation for all \(x_{\text{to}} \in x_{\text{to}_\text{set}}\):
    
    \[ x'_{\text{from}} \mathbf{H} x_{\text{from}} > x'_{\text{to}} \mathbf{H} x_{\text{to}} \]
    
    \item Calculate a list of candidate \(M\) values for each \(x_{\text{to}} \in x_{\text{to}_\text{set}}\) through:
    
    \begin{equation}
    \label{eq:M_est}
    M > \frac{x'_{\text{from}} \mathbf{H} x_{\text{from}} - x'_{\text{to}} \mathbf{H} x_{\text{to}}}{(A x_{\text{to}} - b)^2 - (A x_{\text{from}} - b)^2}
    \end{equation}

    \item Select the largest value from the above list as the low bound of the penalty coefficient \(M\) for the QUBO formulation.

\end{itemize}

\begin{algorithm}[H]
\caption{Penalty Coefficient Estimation}
\label{al:monte_carlo_method}
\begin{algorithmic}[1]
    \STATE \textbf{Input:} Matrix $A$, vector $b$, matrix $\mathbf{H}$, number of iterations $N$
    \STATE \textbf{Output:} Value $M$
    
    \STATE Generate a set of feasible solutions, say $x_{\text{set}}$, using the Monte Carlo method with $N$ iterations
    \STATE Find a feasible solution in $x_{\text{set}}$, say $x_{\text{from}}$, that minimize $Ax - b$
    \STATE From the remaining feasible solutions in $x_{\text{set}}$, find the subset, say $x_{\text{to}_\text{set}}$, such that $x'_{\text{from}} \mathbf{H} x_{\text{from}} > x'_{\text{to}} \mathbf{H} x_{\text{to}}$, for all $x_{\text{to}} \in x_{\text{to}_\text{set}}$

    \FOR{each $x_{\text{to}} \in x_{\text{to\_set}}$}
        \STATE Calculate a candidate $M$ value using Equation~\ref{eq:M_est}.
        \STATE Add it into a set, say $M_L$
    \ENDFOR
    
    \STATE Select the largest element in $M_L$ as the lower bound of the penalty coefficient $M$
\end{algorithmic}
\end{algorithm}

\subsubsection{Validation of Penalty Coefficient Estimation}

The efficient frontier, given by SLSQP (Sequential Least Squares Quadratic Programming) method \cite{kraft1988software}, serves as a comparison basis for validating the proposed estimation method.

\begin{figure}[h]
  \centering
  \includegraphics[width=0.45\textwidth]{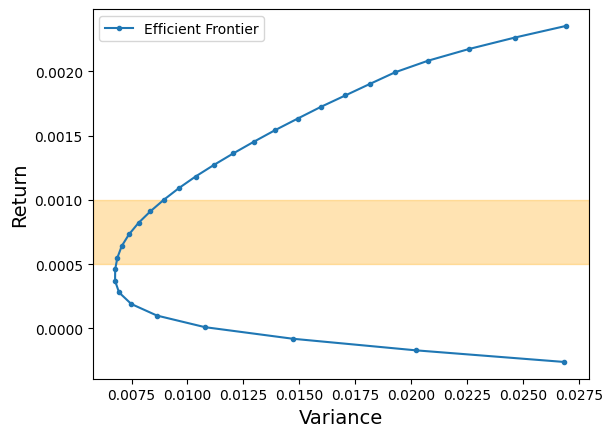} 
  \caption{A selected slicing range for $b$ (The orange shaded region is the sampling region for \(b\); blue line is the efficient frontier)}
  \label{fig:Region}
\end{figure}

\begin{figure}[h]
  \centering
  \includegraphics[width=0.4\textwidth]{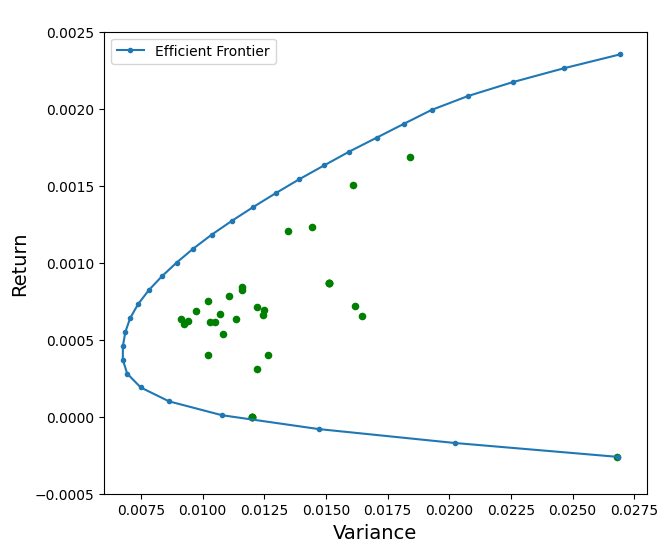} 
  \caption{The results derived from the QUBO formulation (green dots represent the results; blue line is the efficient frontier)}
  \label{fig:Region_res}
\end{figure}

Figure~\ref{fig:Region} and Figure~\ref{fig:Region_res} obtained by using the following steps are used to visually validate the effectiveness of the proposed method for the QUBO problem of interest:

Firstly, solving the problem by alternating the value of \(b\) during the annealing process yields a series of solutions that are fixed points on the efficient frontier.

Secondly, while it is possible to select as many values of $b$ as possible from a broad range of expected returns along the efficient frontier to calculate the estimate of the low bound of $M$ and solve the problem accordingly, it could be impractical due to time or resource constraints. To streamline this approach while ensuring diverse exploration of the efficient frontier, it is a practical way to focus the evaluation on a slicing range of expected returns at a time as shown by the orange shaded region in Figures~\ref{fig:Region}, so as to accelerate the evaluation process while ensuring that certain possible outcomes on the efficient frontier are thoroughly investigated. For instance, all the feasible outcomes  (green dots) relative to the efficient frontier (blue
line) for a slicing range of $b$ are illustrated in Figure \ref{fig:Region_res}.

\subsection{Exploration of Two-Stage Search}

The algorithm of the proposed two-stage search begins with the first stage by solving Equation~\ref{eq:qubo_MPT_normal} with an arbitrarily selected return ($R$) while restricting the values of \(x_i\) to be binary (1 or 0) to obtain a list of feasible solutions, $L_b$, which serves as the starting point for further refinement. This broad (rough) search in the first stage helps quickly obtain an initial result, which is then used as a starting configuration by a refined search to improve the solution accuracy in the second stage.

Subsequently, the algorithm iterates over each element in \( L_b \) to convert the special type of inequality constraint to equivalent quadratic penalty term~\cite{glover2018tutorial}, which is then multiplied by an adjusted penalty coefficient of $M$ to obtained the final quadratic penalty term associated with the two-stage search. 

After completing the adjusted quadratic penalty term, it is then used to transform the constrained problem into a modified QUBO problem corresponding to Equation~\ref{binary_sub}. Two QUBO problems in the form of Equation~\ref{binary_sub} - one using adjusted penalty term (with two-stage search), and the other using original one (without two-stage search) - are then constructed and solved to yield their respective optimal configurations from which the expected returns and variances can be derived for performance evaluation. The procedures of the proposed two-stage search are summarized in Algorithm~\ref{al:result_estimation_strategy}.

\begin{algorithm}
\caption{Two-Stage Search}
\label{al:result_estimation_strategy}
\begin{algorithmic}[1]
    \STATE \textbf{Input:} $\theta$, $R$, $L_b$, $K$, and $M$
    \STATE \textbf{Output:} Updated lists $L_{b\_k\_with}$ and $L_{b\_k\_without}$
    \STATE Set $\theta$ for both Equation~\ref{eq:qubo_MPT_normal}  and Equation~\ref{binary_sub} // Set parameters for both equations
    \STATE Arbitrarily select a return $R$ for Equation~\ref{eq:qubo_MPT_normal}  // Choose a return value randomly
    \STATE Solve Equation~\ref{eq:qubo_MPT_normal}  subject to binary variables to obtain a list of feasible solutions, denoted as $L_b$ // Complete the first stage
    \STATE $temp_{out} \gets 0$ 
    \FOR{each $e \in L_b$  // Convert inequality to quadratic penalty}
        \STATE Initialize $temp \gets 0$
        \FOR{$i \gets 0$ \textbf{to} $K - 1$}
            \STATE $temp \gets temp - x_{i, e}$ // $x_{i,k}$ from Equation~\ref{binary_sub}
            \IF{$i \neq K - 1$}
                \FOR{$\text{cross\_term} \gets i + 1$ \textbf{to} $K - 1$}
                    \STATE $temp \gets temp + x_{i, e}\times x_{\text{cross\_term}, e}$
                \ENDFOR
            \ENDIF
        \ENDFOR
        \STATE $temp \gets temp + 1$ 
        \STATE $temp_{out} \gets \text{random number in } (0, 0.5] \times M \times temp$ // Adjust penalty coefficient
    \ENDFOR
    \STATE Initialize $H_1 \gets$ QUBO problem in the form of Equation~\ref{binary_sub}
    \STATE Initialize $H_2 \gets $ QUBO problem in the form of Equation~\ref{binary_sub} while using the adjusted quadratic penalty term $temp_{out}$
    \STATE Solve $H_1$ and $H_2$ respectively, and collect results for comparison:
    \begin{enumerate}
        \item Solve $H$, then output optimal configuration to calculate the expect return and variance
        \item Solve $H$, then output optimal configuration to calculate the expect return and variance
    \end{enumerate}

\end{algorithmic}
\end{algorithm}

\section{Results}
\subsection{Performance of Parameter Tuning}

For experimental evaluation, a quarterly based dataset consisting of 40 assets in S\&P 500 (Appendix) spanning a decade are collected. At the outset of each quarter, the efficient frontier is established, and Sharpe ratios~\cite{sharpe1998sharpe} are calculated for each portfolio configuration. The experiments are performed with two different strategies for configuration selection during search: 1) choosing the portfolio configuration with the highest three Sharpe ratios for processing in each subsequent quarter. 2) retaining the current portfolio configuration unless the one in the subsequent quarter presents a higher Sharpe ratio, in which case the new configuration is adopted. These strategies are applied with an aim to choose a good result within a quarter and observe whether the frequent change of configuration might have a positive result in the long run.

To investigate the effect of parameter tuning of \(\theta\) and \(M\) on the performance, the analysis begins by exploring the relation between the parameters using Equation \ref{eq:qubo_MPT_normal} and Equation \ref{eq:M_est}. From these equations, a comprehensive range of \(M\) values is derived for various combinations of $K$ and \(\theta\) values, and it is observed that certain ranges of \(M\) values leading to failed results or null output. The subsequent analysis focuses on the effects of different $K$ values on the selection of appropriate $\theta$ values and $M$ values for the QUBO formulation under study. It is observed that in the cases of \(K = 10\) and \(K = 20\), setting the coefficient $\theta$ to \(2^{25}\) or above can cause the hardware solver to fail in finding any valid solutions due to the calculation precision of coefficients of QUBO formulation being limited to, for examples, 76 bits for linear term and 64 bits for quadratic term on the current state-of-the art digital annealer. Consequently, the maximum allowable value of coefficient for $\theta$ in cases is capped at \(2^{22}\). In the case of $K = 20$, $\theta = 2^5$ is precluded as the minimum coefficient because it yields results predominantly consisting of zeros for the QUBO problem represented by Equation~\ref{binary_sub}. Those observations underscore the importance to evaluate further how the variations in \(\theta\) along with the effective value(s) of $M$ can influence the outcomes and thus the performance.

Table~\ref{tab:monte_result_grouped_fixed} lists the value(s) of $M$, in the order of function of $\theta$, associated with various combinations of values of $K$ and $\theta$ that are applicable to Equation~\ref{eq:M_est}, except that the combinations with $K=N/A$ are applicable to Equation~\ref{eq:qubo_MPT_normal} only.

\begin{table}[htbp]
    \centering
    \scriptsize 
    \caption{The value(s) of $M$ for various combinations of values of \(K\) and \(\theta\) (Note: The combinations with $K=N/A$ are applicable to Equation~\ref{eq:qubo_MPT_normal} only)}
    \label{tab:monte_result_grouped_fixed}
    \begin{tabular}{>{\centering\arraybackslash}p{1cm} >{\centering\arraybackslash}p{3cm} >{\raggedright\arraybackslash}p{3cm}}
        \toprule
        \textbf{$K$} & \textbf{\(\theta\)} & \textbf{\(M\) (in the order of function of \(\theta\))} \\
        \midrule
        \multirow{4}{*}{5} 
        & \(2^{10}\) & \(O(\theta^2), O(\theta \log \theta)\) \\
        & \(2^{15}\) & \(O(\theta^2), O(\theta \log \theta), O(\theta)\) \\
        & \(2^{20}\) & \(O(\theta^2), O(\theta \log \theta), O(\theta)\) \\
        & \(2^{22}\) & \(O(\theta^2), O(\theta \log \theta), O(\theta)\) \\
        & \(2^{25}\) & \(O(\theta^2)\) \\
        \midrule
        \multirow{4}{*}{10} 
        & \(2^5\)  & \(O(\theta^2)\) \\
        & \(2^{10}\) & \(O(\theta^2), O(\theta \log \theta)\) \\
        & \(2^{15}\) & \(O(\theta^2), O(\theta \log \theta)\) \\
        & \(2^{20}\) & \(O(\theta^2), O(\theta \log \theta)\) \\
        & \(2^{22}\) & \(O(\theta^2), O(\theta \log \theta)\) \\
        \midrule
        \multirow{4}{*}{20} 
        & \(2^{10}\) & \(O(\theta^2), O(\theta \log \theta)\) \\
        & \(2^{15}\) & \(O(\theta^2), O(\theta \log \theta)\) \\
        & \(2^{20}\) & \(O(\theta^2), O(\theta \log \theta)\) \\
        & \(2^{22}\) & \(O(\theta \log \theta)\) \\
        \midrule
        \multirow{2}{*}{N/A}
        & \(1, 10\)  & \(O(\theta^2)\) \\
        & \(100, 1000, 10000, 100000\) & \(O(\theta^2), O(\theta \log \theta)\) \\
        \bottomrule
    \end{tabular}
\end{table}

\begin{figure}[h]
  \centering
  \includegraphics[width=0.45\textwidth]{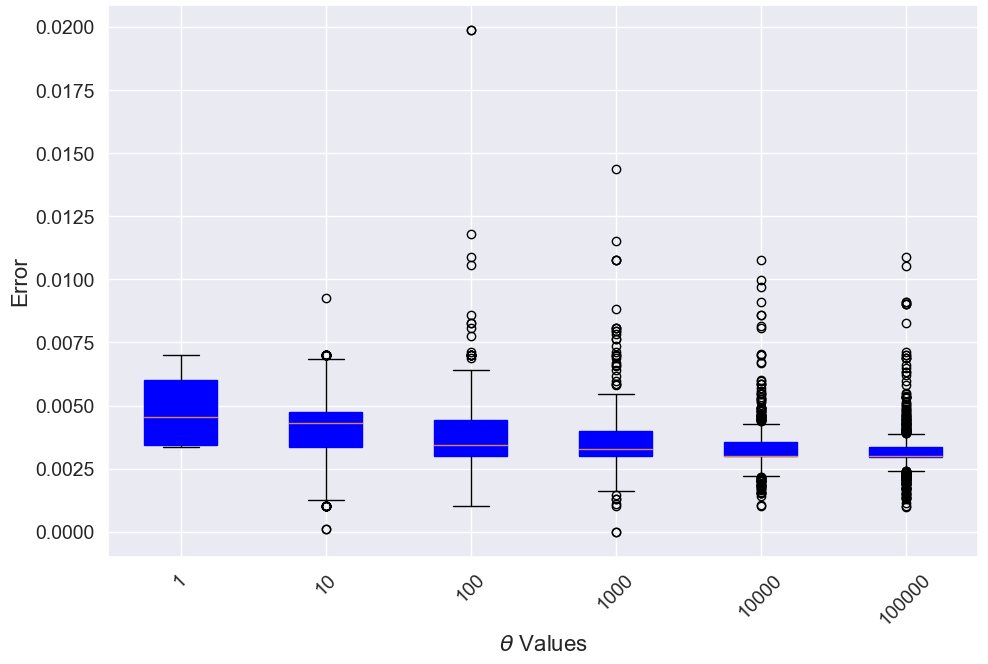}
  \caption{Box-plot of results by coefficient \(\theta\)}
  \label{fig:MPT_normal_all_coeff}
\end{figure}

\begin{figure}[h]
  \centering
  \includegraphics[width=0.45\textwidth]{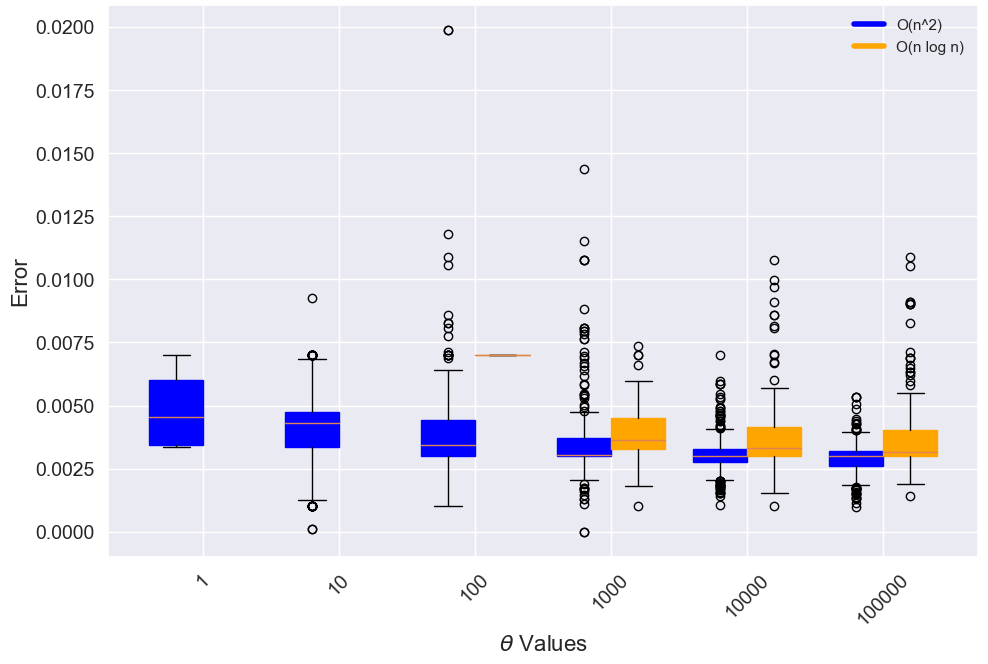}
  \caption{Box-plot by coefficient \(\theta\) with different order of $M$ values}
  \label{fig:MPT_normal_O2}
\end{figure}

As can be seen in Figure~\ref{fig:MPT_normal_all_coeff}, large \(\theta\) values can improve solution accuracy. In Figure~\ref{fig:MPT_normal_O2}, the results are categorized by two different orders of $M$ values to facilitate assessment of solution accuracy with respect to different combinations of parameters' values. It is shown that the solutions obtained by employing higher effective value of $M$ with \(\theta\), particularly at values of 1000, 10000, and 100000, have higher quality than those with lower effective value of $M$, and besides maintain a similar level of quality regardless of different \(\theta\) values. It is noted, however, that the relatively poor solution quality associated with the settings of \(\theta = 100\) and $M$ in the range of \(O(\theta \log \theta)\) highlights the challenges of finding optimal solution due to null results.

\begin{figure}[h]
  \centering
  \includegraphics[width=0.45\textwidth]{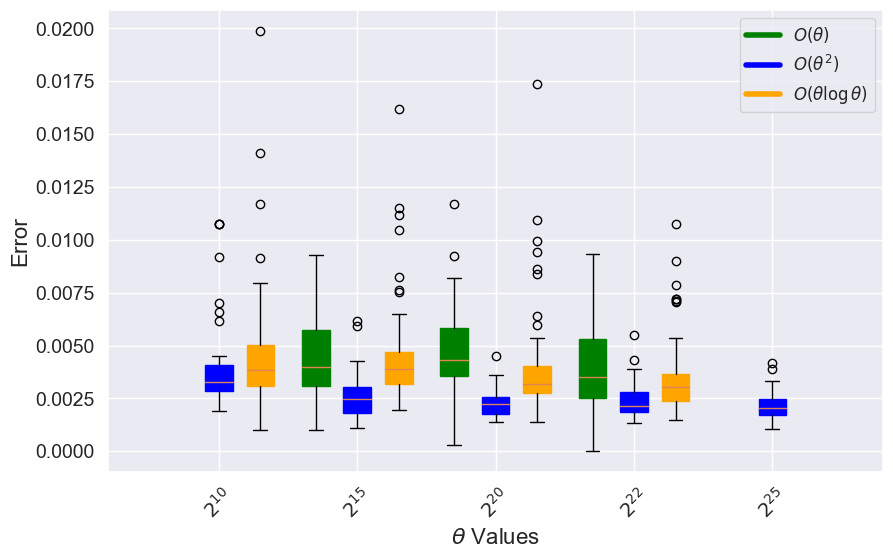}
  \caption{Box-plot by coefficient \(\theta\) with different order of $M$ values for K=5}
  \label{fig:MPT_binary_5_comparison}
\end{figure}

In the case of \( K = 5 \), the error boxes for various \( \theta \) values shows a trend that the error decreases as \( \theta \) value increases. From the error boxes categorized by the orders of $M$ values shown in Figure \ref{fig:MPT_binary_5_comparison}, it is observed that unlike the category of \( O(\theta) \), the box-plots in the categories of \( O(\theta \log \theta) \) and \( O(\theta^2) \) both show a consistent trend in error reduction as \( \theta \) value increases.

\begin{figure}[h]
  \centering
  \includegraphics[width=0.45\textwidth]{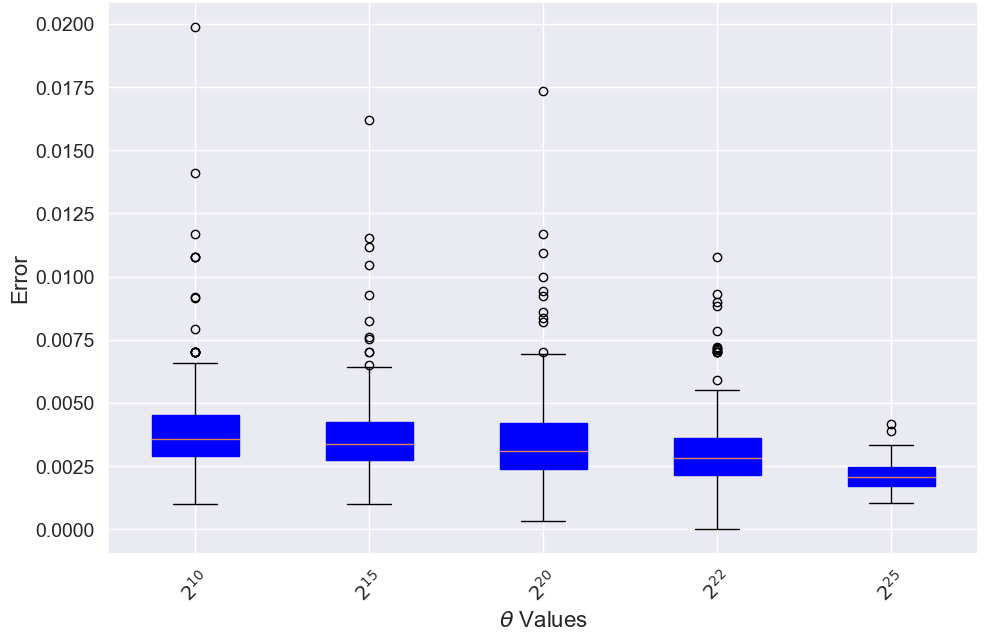}
  \caption{Box-plot by coefficient \(\theta\) for $K=5$}
  \label{fig:MPT_binary_5_all}
\end{figure}

\begin{figure}[h]
  \centering
  \includegraphics[width=0.45\textwidth]{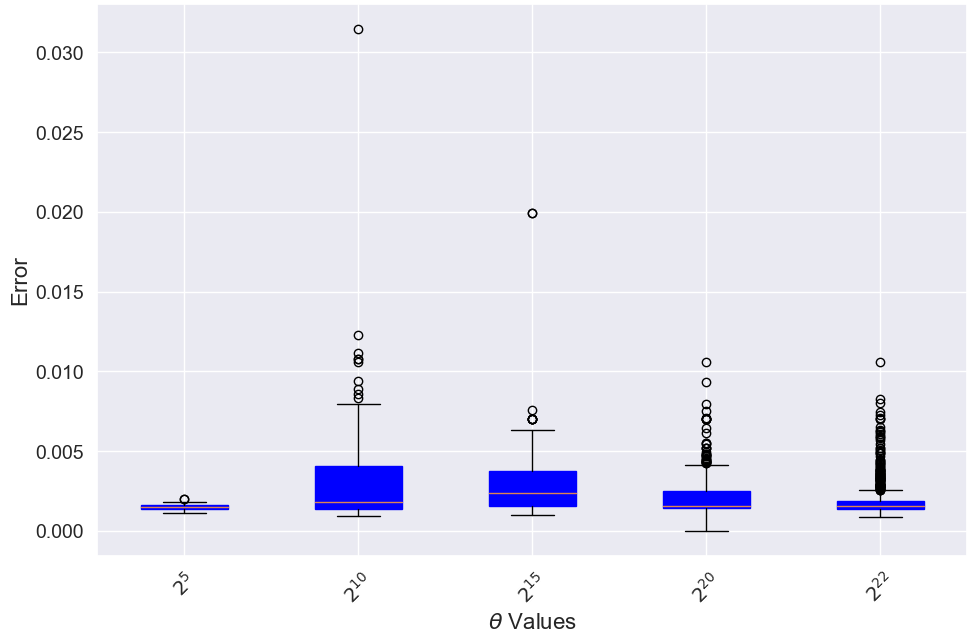}
  \caption{Box-plot by coefficient \(\theta\) fot $K=10$}
  \label{fig:MPT_binary_10_all}
\end{figure}

\begin{figure}[h]
  \centering
  \includegraphics[width=0.45\textwidth]{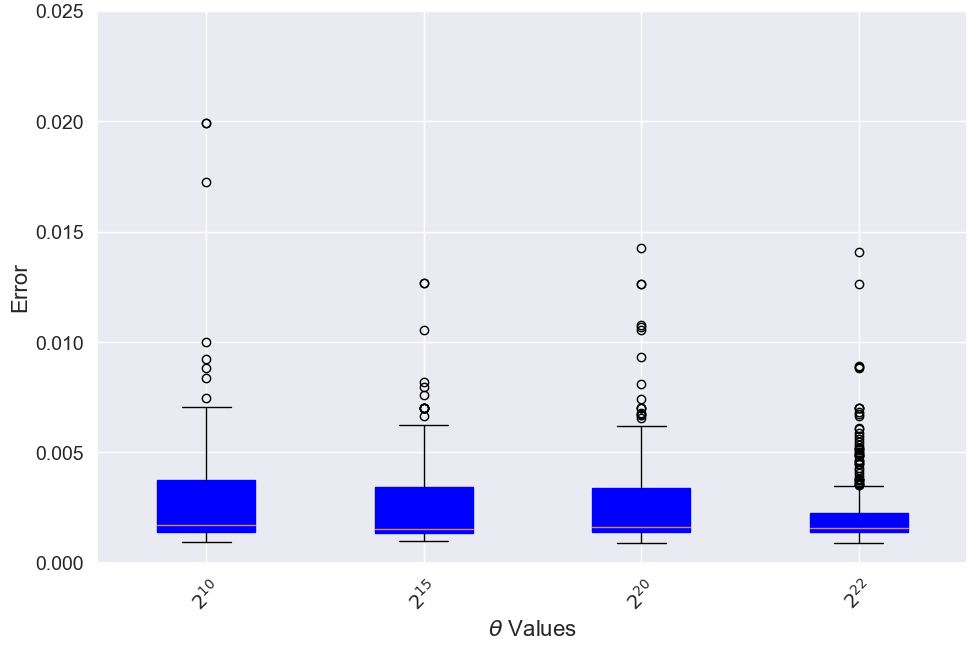}
  \caption{Box-plot by coefficient \(\theta\)  for $K=20$}
  \label{fig:MPT_binary_20_all}
\end{figure}

In the cases of \(K = 10\) and \(K = 20\), it is observed from the box-plots in Figures~\ref{fig:MPT_binary_10_all} and \ref{fig:MPT_binary_20_all} that both of them respectively exhibit a trend of slight error reduction as $\theta$ value increases. Notably, in the case of \( K = 10 \), despite the variability in  the error box for $\theta = 2^5$ is much smaller than those for the other $\theta$ values, most of the search runs in this setting yield null results, which means only a small amount of data are available for evaluation, and therefore care should be taken of using the error box in this case of setting as a measure of performance.

\begin{figure}[h]
  \centering
  \includegraphics[width=0.45\textwidth]{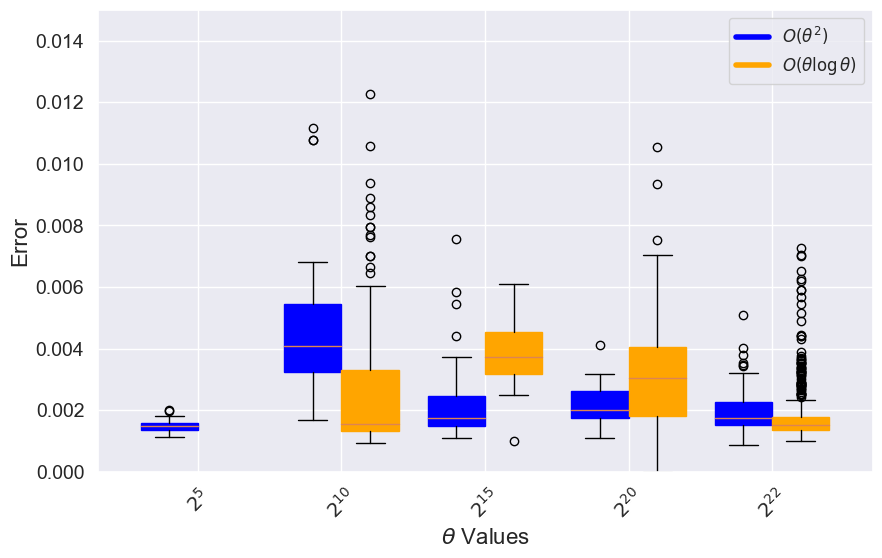}
  \caption{Box-plot by coefficient \(\theta\)  with different order of $M$ values for $K=10$}
  \label{fig:MPT_binary_10_comparison}
\end{figure}

\begin{figure}[h]
  \centering
  \includegraphics[width=0.45\textwidth]{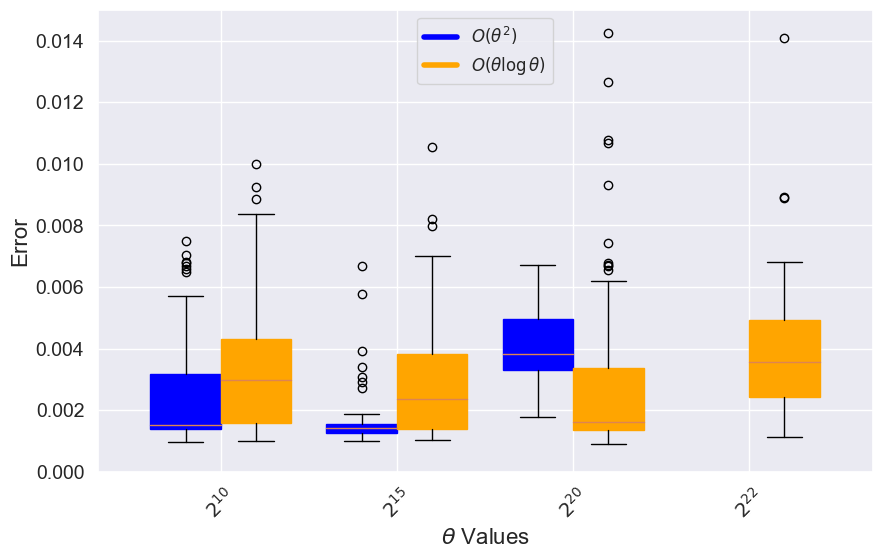}
  \caption{Box-plot by coefficient \(\theta\) with different order of $M$ values for $K=20$}
  \label{fig:MPT_binary_20_comparison}
\end{figure}

It is, however, observed from the box-plots in Figures \ref{fig:MPT_binary_10_comparison} and \ref{fig:MPT_binary_20_comparison} that the distributions of errors associated with different orders of $M$ values do not hold the previous similar trend of error reduction as \( \theta \) value increases. This observation suggests that as the number of binary variables reaches a certain threshold, the distribution of errors becomes random and less predictable among the various \( \theta \) values as a result of using larger $K$ values.

From the analysis of the box-plots in Figures~\ref{fig:MPT_binary_5_all}, ~\ref{fig:MPT_binary_10_all}, and ~\ref{fig:MPT_binary_20_all}, the parameter setting of \(K = 10\) and \(\theta = 2^{22}\) is deemed preferable for experimental evaluation, as is supported by the \(K\) value being not large enough to increase computation time excessively, or small enough to cause imprecision in the results.

\begin{figure}[h]
  \centering
  \includegraphics[width=0.48\textwidth]{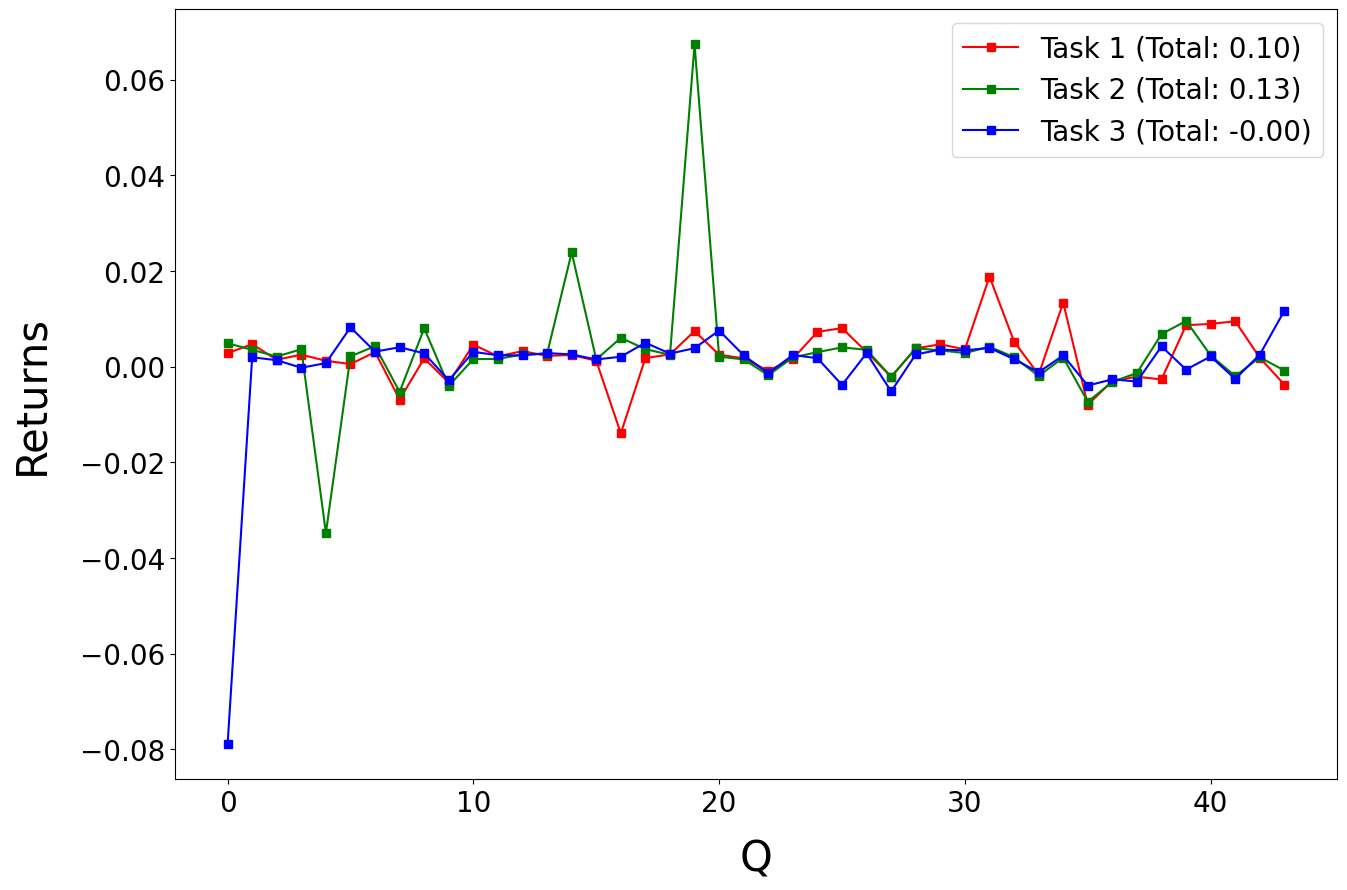}
  \caption{Experimental evaluation for the first strategy (Q represents quarter)}
  \label{fig:Backtesting_f}
\end{figure}

\begin{figure}[h]
  \centering
  \includegraphics[width=0.48\textwidth]{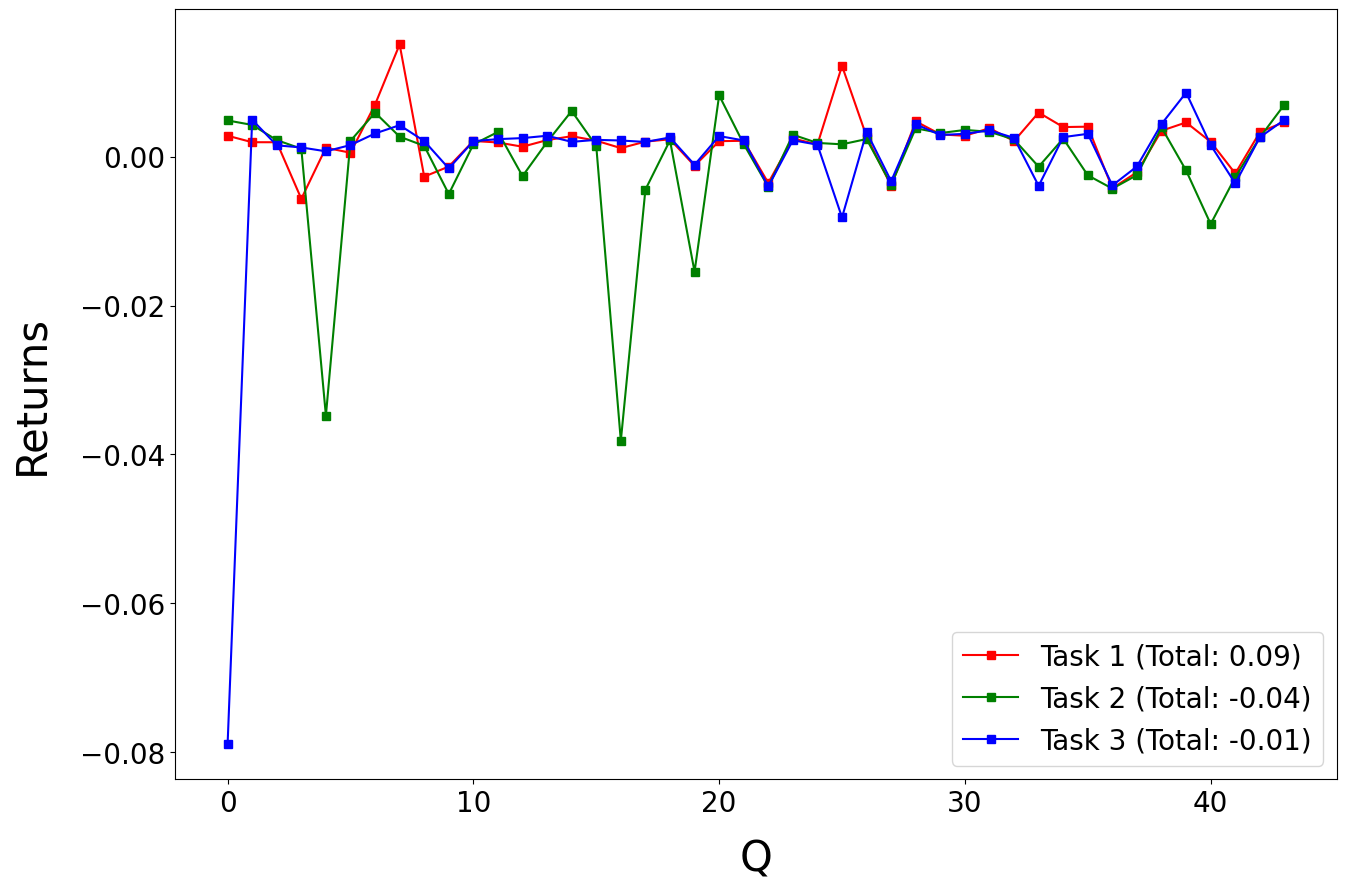}
  \caption{Experimental evaluation for the second strategy (Q represents quarter)}
  \label{fig:Backtesting_s}
\end{figure}

Figures~\ref{fig:Backtesting_f} and~\ref{fig:Backtesting_s} illustrate respectively the experimental results of performance evaluation of the two strategies for configuration selection over three different selection tasks of which Task 1 uses the highest Sharpe ratio value each quarter, and Tasks 2 and 3 use the second and third highest values, respectively. The total return is obtained by summing all the returns from each quarter. Although there are some noticeable differences in returns at the beginning quarters, the first strategy consistently outperforms the second strategy in total return in the long term. This suggests a significant temporal correlation in market trends, where the first strategy effectively captures more advantageous returns by continually adapting to the highest current Sharpe ratio. In contrast, the second strategy updates the portfolio configuration only when the current quarter's Sharpe ratio exceeds the previous one, potentially retaining outdated configurations due to its strategy. As both strategies employing Task 1  have yielded a positive total return, further exploration is intended to assess the impact of various combinations of \(K\) and \(\theta\) values on backtesting performance.

\subsection{Performance of Two-Stage Search}

To facilitate the performance evaluation, the results obtained from the experimental study using Algorithm~\ref{al:result_estimation_strategy} are divided into two categories according to those obtained with the two-stage search and those without for comparison, and the efficient frontier plot and box-plot are used as the tools to visually demonstrate the performance gained by employing the two-stage search method.

\begin{figure}[h]
  \centering
  \includegraphics[width=0.45\textwidth]{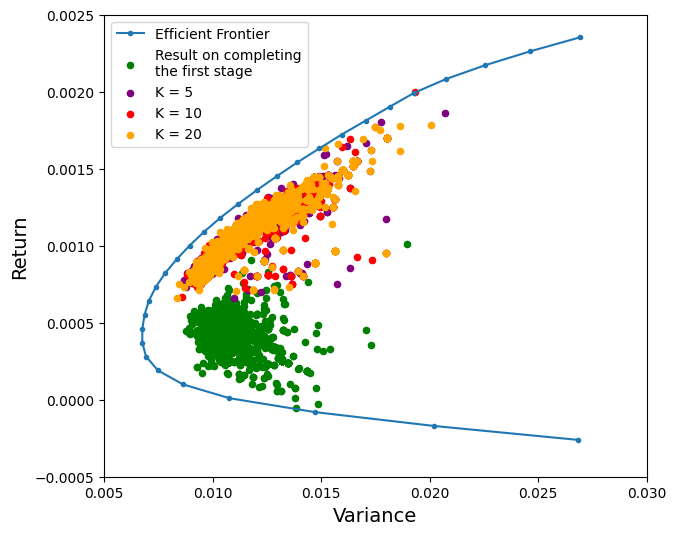}
  \caption{Result of applying two-stage search at different $K$ values in the efficient frontier plot}
  \label{fig:ef_all}
\end{figure}

\begin{figure}
  \centering
  \includegraphics[width=0.45\textwidth]{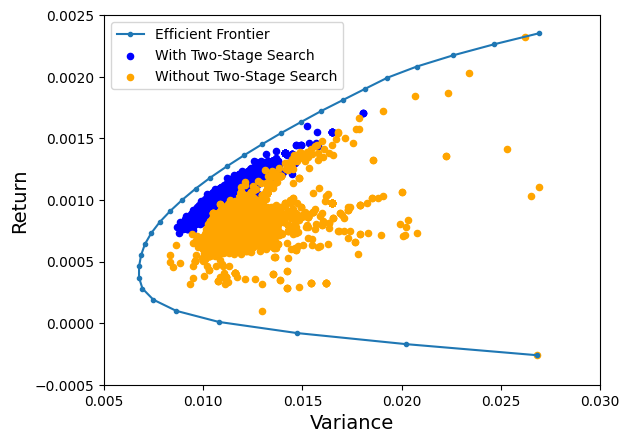}
  \caption{Comparison of results with and without  applying two-stage search for $K = 5$}
  \label{fig:ef_5}
\end{figure}

\begin{figure}
  \centering
  \includegraphics[width=0.45\textwidth]{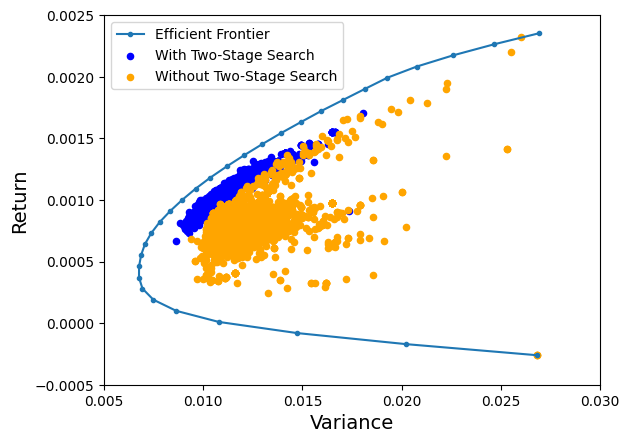}
  \caption{Comparison of results with and without applying two-stage search for $K = 10$}
  \label{fig:ef_10}
\end{figure}

\begin{figure}
  \centering
  \includegraphics[width=0.45\textwidth]{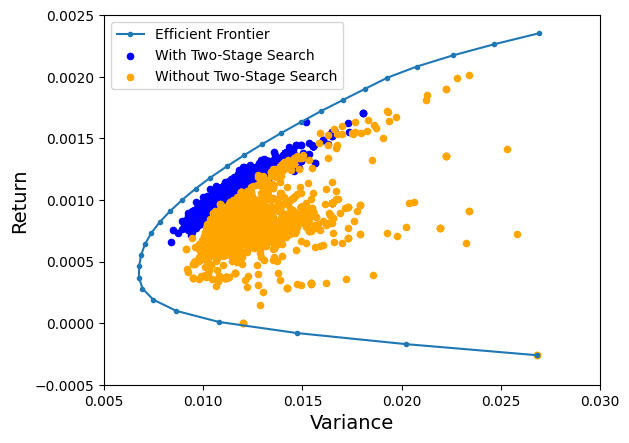}
  \caption{Comparison of results with and without applying two-stage search for $K = 20$}
  \label{fig:ef_20}
\end{figure}

\begin{figure}
  \centering
  \includegraphics[width=0.48\textwidth]{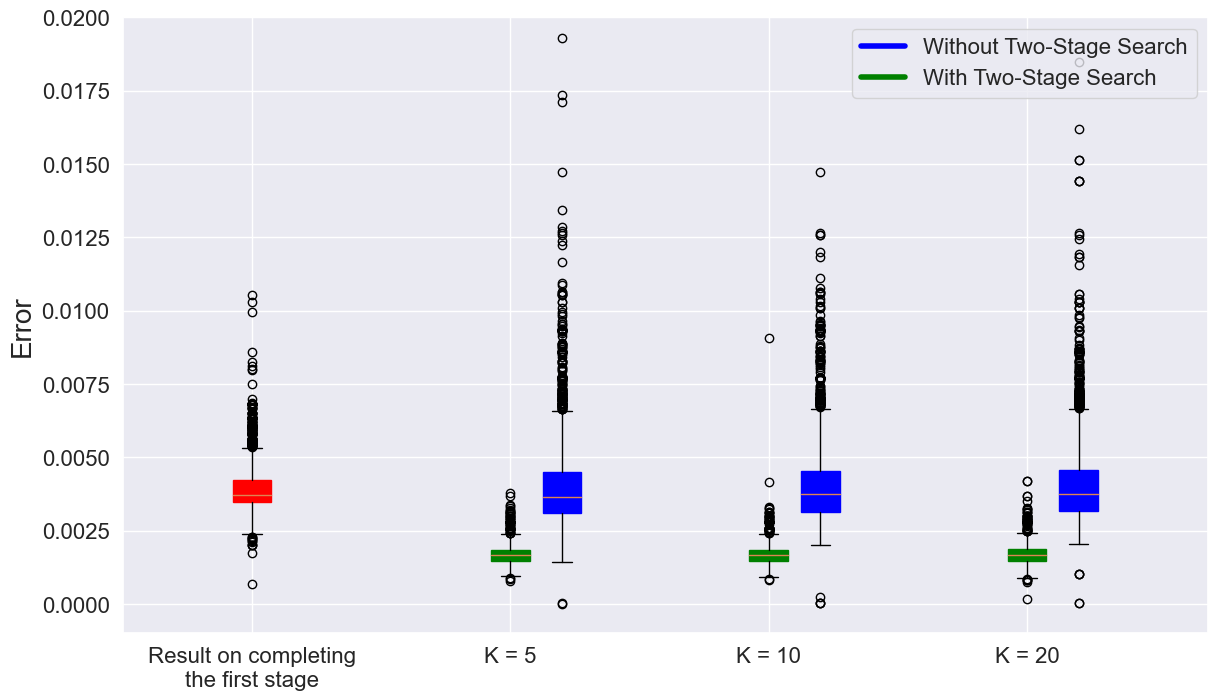}
  \caption{Box-plot of results with and without applying two-stage search for different $K$ values}
  \label{fig:ef_error}
\end{figure}

From Figure~\ref{fig:ef_all}, it is observed that there is a clear performance gap between the results obtained on completing the first stage of the two-stage search (green dots) and those obtained on completing the entire two-stage search (all the other dots), and it is obvious for higher $K$ values to bring about results better aligned with the efficient frontier. Furthermore, from the results shown in the efficient frontier plots of Figures~\ref{fig:ef_5}-\ref{fig:ef_20}, it is observed that, thanks to the first stage search to identify initial configurations, there is a clear performance advantage of using the two-stage search (blue dots) over not using (dispersed orange dots), regardless of different $K$ values. Additionally, the above observations can also be evidenced by Figure~\ref{fig:ef_error}, where both the performance gap between the first stage search and the two-stage search, and the performance advantage of using the two-stage search can be better appreciated through the box-plot of results.

To evaluate the speedup that can be gained by using two-stage search, the computational time improvement ratio for two-stage search, defined as follows, is used: 
\[
\text{computational time improvement ratio: } \frac{T_{\text{without}}}{T_{\text{with}}}
\]
where \(T_{\text{without}}\) and \(T_{\text{with}}\) represent respectively the computational time without and with applying two-stage search. This ratio helps quantify how effective two-stage search is to reduces the computational time.

\begin{figure}
  \centering
  \includegraphics[width=0.45\textwidth]{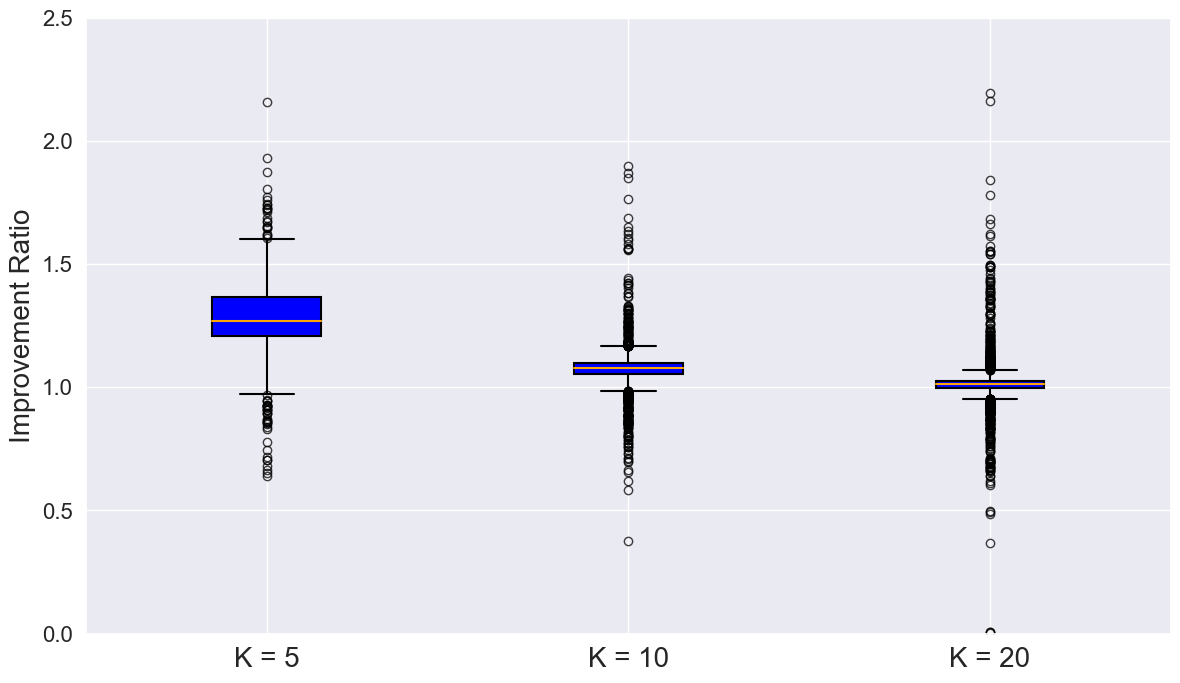}
  \caption{Computational time improvement ratios for different $K$ values}
  \label{fig:annealing_rate}
\end{figure}

Figure~\ref{fig:annealing_rate} illustrates the computational time improvement ratios for various \(K\) values (5, 10, and 20). As can be seen from the box-plot that lower value of $K$ tends to have higher computational time improvement ratio than higher value of $K$ in that a significant speedup is gained at $K=5$, while nearly no speedup is observed at $K=20$. The result suggests that as \(K\) increases to a large value, the  number of variables in the QUBO formulation increases significantly with the introduction of more terms associated with the binary expansion of the decision variables to accommodate proportional investments, which in turn consumes more time  to an extent that the computational time is dominated mainly by the problem size no matter if the two-stage search is applied.

\section{Discussion}

This section will discuss the effects that parameter tuning, encoding granularity, and two-stage search have in shaping the performance of the quantum-inspired portfolio optimization in the QUBO framework.

\subsection{Effect of Parameter Tuning and Encoding Granularity}

The setting of parameter \(\theta\) using the proposed method in QUBO formulations significantly influences the quality of solutions in portfolio management, as evidenced by the analysis of box-plots from Figure~\ref{fig:MPT_normal_all_coeff},~\ref{fig:MPT_binary_5_all},~\ref{fig:MPT_binary_10_all}, and~\ref{fig:MPT_binary_20_all}. Higher \(\theta\) values suggest more reliable outcomes, in terms of lower median errors and less variability, where the effect is particularly pronounced at larger \(\theta\) values. The selection of the penalty coefficient $M$ using Algorithm~\ref{al:monte_carlo_method} has demonstrated its effectiveness in tuning parameter \(\theta\). Furthermore, different settings of the granularity parameter \(K\) at 5, 10, and 20 respectively, are used to assess its impact on computational complexity and solution accuracy. Currently, a preliminary result for choosing a preferable \(K\) has been obtained that can effectively solve the target problem, however, more effective strategy or method for selecting optimal $K$ is pending further investigation.

Moreover, the QUBO formulation with higher orders of $M$ values, such as \(O(\theta^2)\) and \(O(\theta \log \theta)\), can yield more accurate solutions, but also can cause the hardware solver to fail in finding valid solutions due to the calculation precision of coefficients of QUBO formulation being limited by the hardware solver. Therefore, further investigation of  optimal setting specifically in the cases of even higher orders of $M$ values (e.g., \(O(\theta^3)\)) is highly necessary.

\subsection{Effect of Two-Stage Search}

By using the two-stage search in the context of quantum-inspired optimization, the experimental results have shown a better performance, in terms of improved solution accuracy at a cost of only slight increase in computational complexity, in finding the optimal solution. It is observed from Figures~\ref{fig:ef_5}, \ref{fig:ef_10}, and \ref{fig:ef_20} that the performance gaps between with and without applying the two-stage search are significant, however, their respective performances are almost the same visually regardless of different $K$ values, as can also be observed from the box-plots in Figure~\ref{fig:ef_error}. Besides, the computational improvement ratio at the large value of   \(K = 20\) in Figure~\ref{fig:annealing_rate} is so negligibly small that the computational efficiency may be improved without sacrificing the solution accuracy by using smaller $K$ values. Therefore, further investigation is needed into the selection of optimal $K$ along with the parameter tuning strategy to ensure solution accuracy while maintaining fast convergence to solution.

\section{Conclusion}

The quantum-inspired approach in the QUBO framework is promising toward portfolio optimization. The proposed parameter tuning techniques can find a good solution in terms of accuracy and efficiency based on the testing results with a real-world dataset of 40 assets spanning over a ten-year period. The use of the proposed preprocessing technique of two-stage search further strengthens the quantum-inspired optimization framework for portfolio management of which the effectiveness is evidenced by its ability to improve computational efficiency while maintaining solution accuracy with appropriate setting of parameters, making it a potential useful tool for tackling portfolio optimization problems. The future work will focus on extending the methods to the more general setting of QUBO model for complex portfolio optimization problems.

\section*{Acknowledgements}

The authors would like to thank Fujitsu Taiwan Ltd. for partial financial and resource support.

\newpage

\appendix
\section{List of Assets}

The table lists the assets used in the analysis with their corresponding tickers and company names:

\begin{table}[h]
\centering
\begin{tabular}{lll}
\toprule
Ticker & Company Name \\
\midrule
MMM   & 3M \\
AXP   & American Express \\
AMGN  & Amgen \\
AAPL  & Apple Inc. \\
BA    & Boeing \\
CAT   & Caterpillar Inc. \\
CVX   & Chevron Corporation \\
CSCO  & Cisco Systems \\
KO    & The Coca-Cola Company \\
GS    & Goldman Sachs \\
HD    & The Home Depot \\
HON   & Honeywell \\
IBM   & IBM \\
INTC  & Intel \\
JNJ   & Johnson \& Johnson \\
JPM   & JPMorgan Chase \& Co. \\
MCD   & McDonald's \\
MRK   & Merck \& Co. \\
MSFT  & Microsoft \\
NKE   & Nike, Inc. \\
PG    & Procter \& Gamble \\
CRM   & Salesforce \\
TRV   & The Travelers Companies \\
UNH   & UnitedHealth Group \\
V     & Visa Inc. \\
WBA   & Walgreens Boots Alliance \\
WMT   & Walmart \\
DIS   & The Walt Disney Company \\
VZ    & Verizon Communications \\
\bottomrule
\end{tabular}
\label{tab:stocks}
\end{table}

\bibliographystyle{unsrt}
\bibliography{reference}

\begin{thebibliography}{10}

\bibitem{markowits1952portfolio}
Harry~M Markowits.
\newblock Portfolio selection.
\newblock {\em Journal of finance}, 7(1):71--91, 1952.

\bibitem{jin2016constrained}
Yan Jin, Rong Qu, and Jason Atkin.
\newblock Constrained portfolio optimisation: The state-of-the-art markowitz models.
\newblock In {\em International Conference on Operations Research and Enterprise Systems}, volume~2, pages 388--395. Scitepress, 2016.

\bibitem{anagnostopoulos2011mean}
Konstantinos~P Anagnostopoulos and Georgios Mamanis.
\newblock The mean--variance cardinality constrained portfolio optimization problem: An experimental evaluation of five multiobjective evolutionary algorithms.
\newblock {\em Expert Systems with Applications}, 38(11):14208--14217, 2011.

\bibitem{9350582}
Fernando G. D.~C. Ferreira, Amir~H. Gandomi, and Rodrigo T.~N. Cardoso.
\newblock Artificial intelligence applied to stock market trading: A review.
\newblock {\em IEEE Access}, 9:30898--30917, 2021.

\bibitem{chopra2021application}
Ritika Chopra and Gagan~Deep Sharma.
\newblock Application of artificial intelligence in stock market forecasting: a critique, review, and research agenda.
\newblock {\em Journal of risk and financial management}, 14(11):526, 2021.

\bibitem{cohen2022algorithmic}
Gil Cohen.
\newblock Algorithmic trading and financial forecasting using advanced artificial intelligence methodologies.
\newblock {\em Mathematics}, 10(18):3302, 2022.

\bibitem{palmer2021quantum}
Samuel Palmer, Serkan Sahin, Rodrigo Hernandez, Samuel Mugel, and Roman Orus.
\newblock Quantum portfolio optimization with investment bands and target volatility.
\newblock {\em arXiv preprint arXiv:2106.06735}, 2021.

\bibitem{slate2021quantum}
Nicholas Slate, Edric Matwiejew, Samuel Marsh, and Jingbo~B Wang.
\newblock Quantum walk-based portfolio optimisation.
\newblock {\em Quantum}, 5:513, 2021.

\bibitem{10.3389/fphy.2014.00005}
Andrew Lucas.
\newblock Ising formulations of many np problems.
\newblock {\em Frontiers in Physics}, 2, 2014.

\bibitem{grant2021benchmarking}
Erica Grant, Travis~S Humble, and Benjamin Stump.
\newblock Benchmarking quantum annealing controls with portfolio optimization.
\newblock {\em Physical Review Applied}, 15(1):014012, 2021.

\bibitem{lucas2014ising}
Andrew Lucas.
\newblock Ising formulations of many np problems.
\newblock {\em Frontiers in physics}, 2:74887, 2014.

\bibitem{glover2018tutorial}
F~Glover, G~Kochenberger, and Y~Du.
\newblock A tutorial on formulating and using qubo models. arxiv.
\newblock {\em arXiv preprint arXiv:1811.11538}, 2018.

\bibitem{aramon2019physics}
Maliheh Aramon, Gili Rosenberg, Elisabetta Valiante, Toshiyuki Miyazawa, Hirotaka Tamura, and Helmut~G Katzgraber.
\newblock Physics-inspired optimization for quadratic unconstrained problems using a digital annealer.
\newblock {\em Frontiers in Physics}, 7:48, 2019.

\bibitem{rahman2019ising}
Muhammed~Tahsin Rahman, Shuo Han, Navid Tadayon, and Shahrokh Valaee.
\newblock Ising model formulation of outlier rejection, with application in wifi based positioning.
\newblock In {\em ICASSP 2019-2019 IEEE International Conference on Acoustics, Speech and Signal Processing (ICASSP)}, pages 4405--4409. IEEE, 2019.

\bibitem{cohen2020ising}
Eldan Cohen, Avradip Mandal, Hayato Ushijima-Mwesigwa, and Arnab Roy.
\newblock Ising-based consensus clustering on specialized hardware.
\newblock In {\em Advances in Intelligent Data Analysis XVIII: 18th International Symposium on Intelligent Data Analysis, IDA 2020, Konstanz, Germany, April 27--29, 2020, Proceedings 18}, pages 106--118. Springer, 2020.

\bibitem{kalehbasti2021ising}
Pouya~Rezazadeh Kalehbasti, Hayato Ushijima-Mwesigwa, Avradip Mandal, and Indradeep Ghosh.
\newblock Ising-based louvain method: clustering large graphs with specialized hardware.
\newblock In {\em Advances in Intelligent Data Analysis XIX: 19th International Symposium on Intelligent Data Analysis, IDA 2021, Porto, Portugal, April 26--28, 2021, Proceedings 19}, pages 350--361. Springer, 2021.

\bibitem{garcia2022exact}
Marcos~Diez Garc{\'\i}a, Mayowa Ayodele, and Alberto Moraglio.
\newblock Exact and sequential penalty weights in quadratic unconstrained binary optimisation with a digital annealer.
\newblock In {\em Proceedings of the Genetic and Evolutionary Computation Conference Companion}, pages 184--187, 2022.

\bibitem{VERMA2022100594}
Amit Verma and Mark Lewis.
\newblock Penalty and partitioning techniques to improve performance of qubo solvers.
\newblock {\em Discrete Optimization}, 44:100594, 2022.
\newblock Quadratic Combinatorial Optimization Problems.

\bibitem{verma2021variable}
Amit Verma and Mark Lewis.
\newblock Variable reduction for quadratic unconstrained binary optimization.
\newblock {\em arXiv preprint arXiv:2105.07032}, 2021.

\bibitem{sakuler2023real}
Wolfgang Sakuler, Johannes~M. Oberreuter, Riccardo Aiolfi, Luca Asproni, Branislav Roman, and Jürgen Schiefer.
\newblock A real world test of portfolio optimization with quantum annealing, 2023.

\bibitem{app122312288}
Jonas Lang, Sebastian Zielinski, and Sebastian Feld.
\newblock Strategic portfolio optimization using simulated, digital, and quantum annealing.
\newblock {\em Applied Sciences}, 12(23), 2022.

\bibitem{kraft1988software}
Dieter Kraft.
\newblock A software package for sequential quadratic programming.
\newblock {\em Forschungsbericht- Deutsche Forschungs- und Versuchsanstalt fur Luft- und Raumfahrt}, 1988.

\bibitem{sharpe1998sharpe}
William~F Sharpe.
\newblock The sharpe ratio.
\newblock {\em Streetwise--the Best of the Journal of Portfolio Management}, 3:169--185, 1998.

\end{thebibliography}

\end{document}